\documentclass[12pt]{article}
\usepackage{graphicx,amssymb}
\usepackage{ifpdf}
\usepackage{xspace}
\usepackage{multirow}
\usepackage{dblfloatfix}
\usepackage{pdflscape}
\usepackage{setspace}
\usepackage{authblk}
\usepackage{natbib}
\usepackage{enumitem}
\usepackage{hyperref}
\setcitestyle{numbers,open={[},close={]}}

\makeatletter

\usepackage{xcolor}
\definecolor{darkgreen}{HTML}{000000}
\definecolor{darkblue}{HTML}{000000}
\newcommand{\marked}[1]{{\color{darkgreen}#1}}
\newcommand{\blue}[1]{{\color{darkblue}#1}}

\graphicspath{{jwsGraphics/}}
\usepackage[center]{subfigure}

\newcommand{\chao}{ChAO\xspace}
\newcommand{\pro}{Prot\'eg\'e\xspace}
\newcommand{\cprot}{Collaborative Prot\'eg\'e\xspace}
\newcommand{\wpro}{WebProt\'eg\'e\xspace}

\newcommand{\plotwidth}{0.9}

\begin{document}

\thispagestyle{empty} \newpage 

\title{How to Apply Markov Chains for Modeling Sequential Edit Patterns in Collaborative Ontology-Engineering Projects}

\author[1]{Simon Walk}
\author[2]{Philipp Singer}
\author[2,3]{Markus Strohmaier}
\author[4]{Denis Helic}
\author[5]{Natalya F. Noy}
\author[5]{Mark A. Musen}

\affil[1]{IICM, Graz University of Technology, Austria}
\affil[2]{GESIS, Germany}
\affil[3]{University of Koblenz-Landau, Germany}
\affil[4]{KTI, Graz University of Technology, Austria}
\affil[5]{BMIR, Stanford University, USA}

\medskip
\normalsize
\maketitle
\begin{abstract} 
With the growing popularity of large-scale collaborative ontology-engineering projects, such as the creation of the 11$^{\textnormal{th}}$ revision of the International Classification of Diseases, we need new methods and insights to help project- and community-managers to cope with the constantly growing complexity of such projects. 
In this paper, we present a novel application of Markov chains to model sequential usage patterns that can be found in the change-logs of collaborative ontology-engineering projects.
We provide a detailed presentation of the analysis process, describing all the required steps that are necessary to apply and determine the best fitting Markov chain model. 
Amongst others, the model and results allow us to identify structural properties and regularities as well as predict future actions based on usage sequences. We are specifically interested in determining the appropriate Markov chain orders which postulate on how many previous actions future ones depend on. 
To demonstrate the practical usefulness of the extracted Markov chains we conduct sequential pattern analyses on a large-scale collaborative ontology-engineering dataset, the International Classification of Diseases in its $11^{\textnormal{th}}$ revision. To further expand on the usefulness of the presented analysis, we show that the collected sequential patterns provide potentially actionable information for user-interface designers, ontology-engineering tool developers and project-managers to monitor, coordinate and dynamically adapt to the natural development processes that occur when collaboratively engineering an ontology. 
We hope that presented work will spur a new line of ontology-development tools, evaluation-techniques and new insights, further taking the interactive nature of the collaborative ontology-engineering process into consideration. 
\end{abstract}

\section{Introduction}
\label{intro}

In recent years, we have seen significant increase in the use of structured data. In many cases, workers have used ontologies to integrate and interpret this data. As a result, we have seen an increase in the number of large-scale projects, focusing on collaboratively engineering ontologies.
For example, the World Health Organization (WHO) is leading the collaborative online development of the new revision of the International Classification of Diseases (ICD), which represents an important classification scheme that is used in many countries around the world for health statistics, insurance billing, epidemiology, and so on. 
Wikidata\footnote{http://www.wikidata.org}, another collaborative ontology-engineering project initiated by the Wikimedia Foundation,\footnote{http://wikimediafoundation.org} is gathering structured data in multiple languages to link to and between Wikipedia and its different language editions. 
To understand and support the new requirements that this collaborative approach introduces, researchers have analyzed and developed new ontology-engineering tools, such as \cprot and \wpro~\cite{CollaborativeProtege.ISWC, WebProtege.SWJ}. These tools not only provide a collaborative environment to engineer ontologies, but also include mechanisms that are targeted towards augmenting collaboration and increasing the overall quality of the resulting ontologies by supporting contributors in reaching consensus. 
For user-interface designers, community managers as well as project administrators, analyzing and understanding the ongoing processes of how ontologies are engineered collaboratively is crucial. When provided with detailed and quantifiable insights, the used ontology-engineering tools or even the development strategy can be automatically revised and adjusted accordingly. 
Engineering an ontology by itself already represents a complex task; this task becomes even more complex when adding a layer of social interactions on top of the development process. In the light of these challenges, we need new methods and techniques to better understand and measure the social dynamics and processes of collaborative ontology-engineering efforts.

In this work, we want to focus on sequences of actions that users perform when collaboratively engineering ontologies. For example, when the change of a property by a user is succeeded by another change of a property by that user, the two changes can be used to represent the sequence of properties that this specific user has been working on.
Better understanding such sequential processes can help system designers to  increase the quality of an ontology or contributor satisfaction, among other things. To come back to our previous example, if we better understand the process  of how users sequentially edit properties of concepts, we can recommend to users the property that they potentially might want to edit next. Alternatively, we can steer users away from their typical behavior in order to  cover niche parts of the ontology. We know from previous studies, that sequential patterns of human actions can usually be predicted quite well. For example, \citet{song} showed that human mobility patterns are predictable; they also hypothesize that all human activities contain certain regularities that can be detected. We explore whether these regularities might also apply to our ontology-editing sequences. 

Consequently, our main goal in this paper is the presentation of methods and techniques for acquiring detailed insights into these ongoing (sequential) processes when users collaboratively engineer an ontology. Hence, we introduce a novel application of a methodology based on Markov chains. 
We base our elaboration of this method on previous work that has focused on studying human navigational paths through websites \cite{singer_mc}.
We focus not only on the structure of given paths (e.g., the identification of common sequences), but also on the detection of memory (e.g., on how many previous changed properties does the next property a user changes depend on).
We lay our focus on determining the appropriate Markov chain orders which allows us to get insights into on how many previous actions users reason their future actions.  

The \textbf{main objectives} of this paper are:
\begin{itemize}

\item The presentation of a novel application of Markov chains on the change logs of collaborative ontology-engineering projects to gather new insights into the processes that occur when users collaboratively create an ontology.

\item The demonstration of the utility of the presented and adapted Markov chain framework by applying it on a large scale collaborative ontology-engineering project.

\end{itemize}

Tackling these two objectives enables us to answer questions that are of practical relevance for the  development of collaborative ontology-engineering tools, such as: Do users have to  switch frequently between the user-interface sections when working on the ontology? Which concept is a user likely to change next, the one closer  to or further away from the root concept of the ontology? Which change type is a user most likely to perform next? Do users move along the ontological hierarchy when changing content? Can we identify edit behaviors, such as \emph{top-down} or \emph{bottom-up} editing? Do users only reason their future actions on the current ones or do they depend on a series of preceding ones? However, other kinds of questions are conceivable and can be studied in straight-forward manner by researchers by focusing on the methodological aspects presented in this work.

\textbf{Results:} Our results indicate that the application of Markov chains on the change-logs of collaborative ontology-engineering projects provides new and potentially actionable insights into the processes that occur when users collaboratively create an ontology for project administrators and ontology-engineering tool developers.

\textbf{Contributions:} We provide (i) a detailed description of the process for applying Markov chains on the change-logs of collaborative ontology-engineering projects and (ii) an evaluation of the extracted Markov chain models by applying the methodology on the change-logs of ICD-11, representing a large-scale collaborative ontology-engineering project that exhibits Markov chains of varying orders. 
Our \textbf{high-level contribution} is the presentation of a novel approach that can be used to gather new insights into ongoing processes when collaboratively engineering an ontology by making use of Markov chains to model sequential usage sequences. Amongst others, this allows practitioners to identify structural properties and regularities as well as predict future actions based on usage sequences.

The remainder of the paper is structured as follows: In section~\ref{sub:collaborative ontology-engineering} we provide a brief introduction into collaborative ontology-engineering. We then continue to review related work in section~\ref{related work}. In section~\ref{sec:datasets}, we briefly describe and characterize the history of ICD-11 as well as the dataset and the underlying change-log. We continue with the description of the process in section~\ref{the process}, describing all necessary steps to  extract and interpret Markov chains for a given dataset. In section~\ref{evaluation}, we apply the previously described process to ICD-11, extracting Markov chains of different orders for two different types of analyses. In section~\ref{discussion}, we discuss potential implications and conclude our work in section~\ref{conclusion}.

\section{Collaborative Ontology Engineering}
\label{sub:collaborative ontology-engineering}
According to \citet{Gruber-A-1993,borst1997ceo} and \citet{studer1998}, an ontology is an explicit specification of a shared conceptualization. In particular, this definition refers to a machine-readable construct (the formalization) that represents an abstraction of the real world (the shared conceptualization), which is especially important in the field of computer science as it allows a computer (among other things) to ``understand'' relationships between entities and objects that are modeled in an ontology.

The field of collaborative ontology engineering and its environment pose a new field of research with many new problems, risks and challenges. In general, contributors of collaborative ontology-engineering projects, similar to other collaborative online production systems (e.g., Wikipedia), engage remotely (e.g., via the internet or a client--server architecture) in the development process to create and maintain an ontology. \marked{Given the complexity assigned to engineering an ontology, researchers and practitioners have already discussed and proposed different development methodologies. Analogously to the plethora of different software development processes and methodologies (i.e., the Waterfall-Model, agile development or SCRUM), methodologies and guidelines exist for (collaboratively) creating an ontology which define multiple different aspects of the engineering process. For example, the \emph{Human-centered ontology engineering methodology} (HCOME) \cite{kotis2005hcome,kotis2006human,kotis2008supporting} represents such an approach that sets its focus on (continuously and) actively integrating the knowledge worker\blue{---the users who will rely on and use the created ontology---}in the ontology life-cycle (i.e., by including the users in all planning stages, discussions, requirements analyses, etc.). Similarly, the \emph{DILIGENT} process \cite{pinto2004ontoedit, tempich2005argumentation, pinto2009ontology} defines principles for the distributed development of an ontology, including different stakeholders (e.g., developers or users of the ontology, who both have different purposes and needs for the resulting ontology). \citet{debruyne2010gospl, debruyne2012gospl}, proposed the \emph{Grounding Ontologies with Social Processes and Natural Language} (GOSPL) approach and tool in $2010$. Again, a strong focus was put on the formalization of social processes, which directly result in and impact the evolution of the collaboratively engineered ontology.}

\section{Related Work}
\label{related work}

For the analysis and evaluation conducted in this paper, we identified relevant information and publications in the domains of \marked{(i) sequential pattern mining}, (ii) Markov chain models and (iii) collaborative authoring systems. We discuss each domain next.

\begin{figure*}[th]
\centering
{\includegraphics[width=\linewidth]{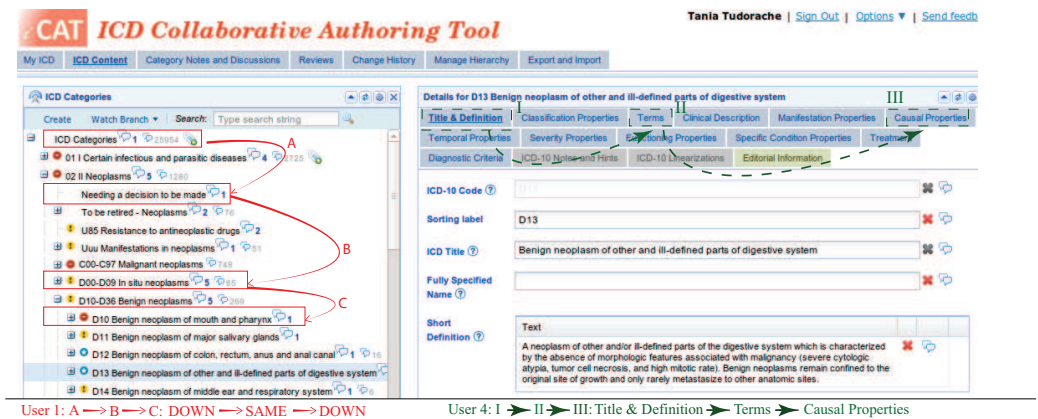}}
\caption{\textbf{The iCAT User-Interface.} A screenshot of the iCAT interface, a custom tailored version of \wpro, developed for the collaborative engineering of ICD-11. The inline annotations represent exemplary transitions between states for two of our three analyses. The letters $A-C$ represent the sequential \emph{Edit-Strategy Path} (see section~\ref{sub:edit strategy paths}) for \textbf{one} user, while the roman numbers $I-III$ constitute a representative sequential path for the \emph{User-Interface Sections Path} analyses (see section~\ref{sub:user-interface tab paths}) for another users. 
Note that for the \emph{Edit-Strategy Paths}, every letter represents the transition between two consecutively changed concepts by the corresponding user. Analogously, for the \emph{User-Interface Sections Paths} each number represents one section of the user-interface that was used by the corresponding users to contribute to the ontology.}
\label{fig:icat}
\end{figure*}

\subsection{Collaborative Authoring Systems}
\label{sub:collaborative authoring systems}

Research on collaborative authoring systems such as Wikipedia has in part focused on developing methods and studying factors that improve article quality or increase user participation. 
For example, \citet{kittur2007bourgeoisie} have shown that for Wikipedia and del.ico.us, two collaborative online authoring systems, participation across users during the initial starting phase is unevenly distributed, resulting in few users (administrators) with a very high participation and contribution rate while the rest of the users (common users) exhibit little if any participation and contributions. However, over time, contributions shift from administrators towards an increasing number of common users, which at the same time still make little contributions individually. Thus, an analysis of the distribution of work across users and articles (as mentioned in \citet{Kittur2008}) can provide meaningful insights into the dynamic aspects of the engineering process.
This line of work is also related to research on problems that are common in these types of environments, such as the \textit{free-riding} and \textit{ramp-up} problems \cite{cabrera2002}. The free-riding problem characterizes the fact that users would rather tend to enjoy a resource than contribute to it. The \textit{ramp-up} problem describes the issue of motivating users to contribute to a system when either content or activity (or both) in the overall system is very low. Researchers have proposed different types of solutions to these---sometimes called---knowledge-sharing dilemmas \cite{cabrera2002}. 
\citet{wilkinson2007cooperation} have shown that the quality of Wikipedia articles correlates with the number of changes performed on these articles by distinct users. 
More recent research which uses collaborative authoring systems, such as Wikipedia as a data source, focuses not only on describing and defining the act of collaboration amongst strangers and uncertain situations that contribute to a digital good \cite{conf/wikis/KeeganGC11} but also on antagonism and sabotage of said systems \cite{shachaf2010beyond}. 
It has also been discovered that Wikipedia editors are slowly but steadily declining \cite{Suh09singularity}. Therefore \citet{conf/wikis/HalfakerKR11} have analyzed what impact reverts have on new editors of Wikipedia, showing that users have a much higher tendency to either stop working on Wikipedia articles after their contributions have been reverted or drastically decrease the amount of contributions.

\marked{Further, \citet{viegas2007talk} have shown that the history of an article and discussion pages in Wikipedia contain valuable information for administrators and moderators. In \cite{viegas2007hidden} the authors conclude that collectives in Wikipedia follow their self-imposed rules regarding well defined and formalized processes, such as featured articles.
\citet{DBLP:journals/semweb/SchneiderGP13, DBLP:conf/cscw/SchneiderSPD13, DBLP:conf/comma/SchneiderPGB10, DBLP:conf/wikis/SchneiderPD12} discussed multiple different aspects and the importance of consensus finding on Wikipedia and the Social Semantic Web, by analyzing the history of articles in said systems, further strengthening the need for tools and analyses to be able to better understand and support digital collaborative endeavors.
}

\subsection{Collaborative Ontology-Engineering Tools}
A number of tools, such as the OntoWiki \cite{Ontowiki.ISWC}, the MoKi \cite{ghidini2009moki}, Soboleo \cite{Soboleo:CKC} or PoolParty \\\cite{schandl2010poolparty} support collaborative ontology engineering, focusing on supporting and augmenting different aspects of  collaborative development processes of ontologies. For example, Semantic MediaWikis \cite{krotzsch2006sm} add semantic capabilities to traditional Wiki systems. They are intended to help users navigating the Wikis by introducing more meaningful semantic links and support of richer queries. Some of the Semantic Wikis available today focus on \textit{enhancing content} with semantic links in order to allow more meaningful navigation and to support richer queries. Semantic Wikis usually associate a page to a particular instance in the ontology, and the semantic annotations are converted into properties of that instance.
As an ontology represents a formalized and abstract version of a specific domain, disagreements between authors on certain subjects can occur. Similar to face-to-face meetings, these collaborative ontology-engineering projects need tools that augment collaboration and help contributors in reaching consensus especially when modeling (controversial) topics of the real world.

In fact, the majority of the literature about collaborative ontology engineering sets its focus on surveying, finding and defining requirements for the tools used in these projects \cite{conf/aaaiss/NoyT08, Groza:2013:CSA:2435464.2435766}.

\pro, and its extensions for collaborative development, such as \wpro and iCAT~\cite{WebProtege.SWJ} (see Figure~\ref{fig:icat} for a screenshot of the iCAT ontology-editor interface) are  prominent  tools that are used by a large community worldwide to develop ontologies in a variety of different projects. Both \wpro and \cprot provide a robust and scalable environment for collaboration and are used in several large-scale projects, including the development of ICD-11~\cite{ICD.ISWC}.

P\"oschko et al. \cite{poeschko-aaai12}, and Walk et al. \cite{walk-ijswis} have created and further developed \emph{PragmatiX}, a tool to browse an ontology and visualize aspects of its history.  PragmatiX also provides quantitative insights into the creation process. The authors applied it to the analysis of the \mbox{ICD-11} project. 

\subsection{Collaborative Ontology-Engineering Analyses}

Strohmaier et al. \cite{j.websem333} investigated the hidden social dynamics that take place in collaborative ontology-engineering projects from the biomedical domain and provided new metrics to quantify various aspects of the collaborative engineering processes. 
\citet{conf/kcap/FalconerTN11} investigated the change-logs of collaborative ontology-engineering projects, showing that contributors exhibit specific roles, which can be used to group and classify these users, when contributing to the ontology.
\citet{10.1371/journal.pcbi.1002630} investigated if the location and specific structural features can be used to determine if and where the next change is going to occur in the Gene Ontology\footnote{http://www.geneontology.org}.
\citet{wang2013analysis} have used association-rule mining  to analyze user editing patterns in collaborative ontology-engineering projects. The approach presented in this paper uses Markov chains to extract much higher detailed user-interaction patterns incorporating a variable number of historic editing information. 

\citet{walk_jbi2014} provided a detailed analysis of the commonalities and differences between five different collaborative ontology-engineering projects. Contrary to the presentation of the Markov chain framework in this paper, \citet{walk_jbi2014} concentrated their efforts on the interpretation of the differences and commonalities in first-order sequential patterns between five different collaborative ontology-engineering projects using aspects of the Markov chain framework presented in detail in this paper.

\marked{\citet{debruyne2013using} presented a generic reputation framework to identify leaders in collaborative ontology-engineering projects. In their framework, they classified users as leaders according to a set of different characteristics (or reputation sensors), such as activity, engagement quality as well as features of the social interaction graph. In \citet{de2009towards}, the authors suggested the use of social performance indicators to gather insights and broaden our understanding of the (ever changing) social arrangement collaboratively evolving an ontology.  

Recently, \citet{van2014method} analyzed behavior-based user profiles in collaborative ontology-engineering projects, relying on GOSPL (Grounding Ontologies with Social Processes and Natural Language) and K-means clustering to group similar users. \citet{di2014evaluating} investigated multiple different features of wiki collaborative features for ontology authoring and showed their impact on the ontology lifecycle and the engineered ontology entities.
}

\marked{
\subsection{Sequential Pattern Mining}
\label{subsec:sequential pattern mining}

\citet{Agrawal:1995:MSP:645480.655281} first addressed the problem of sequential pattern mining in 1995. In their work the authors defined sequential pattern mining as: given a collection of chronologically ordered sequences, sequential pattern mining is about discovering all sequential (chronologically ordered) patterns, weighted according to the number of sequences that contain these patterns. They also introduced AprioriAll and AprioriScale, which also represent the first \emph{a priori} sequential pattern mining algorithm. 
One year later, in 1996, \citet{Srikant:1996:MSP:645337.650382} further included time-constraints and sliding windows to the definition of sequential patterns and introduced the widely popular and used generalized sequential pattern algorithm (GSP).
With this work the authors showed that specific patterns cannot occur more frequently (above a threshold) if a sub-pattern of this pattern occurs less often (below that threshold). Many additional examples of a priori algorithms have been reviewed and discussed in literature \cite{Mannila:1997:DFE:593416.593449, Wang:1994:CPD:191839.191863, Bettini:1996:TCT:237661.237680}, with SPADE \cite{Zaki01spade:an} being one of the most prominently used and referred to algorithms.
One major problem assigned to the a priori based sequential pattern mining algorithms was (in the worst case) the exponential number of candidate generation. As a priori based sequential pattern mining algorithms create (in the worst case) an exponential number of candidates \citet{Han:2000:MFP:342009.335372, pei2001prefixspan} invented so called pattern-growth approaches. They circumvent the exponential candidate generation by strategically expanding found patterns and ignoring patterns that are not present in the data. 

Today, many researchers have adapted different sequential pattern mining algorithms and approaches for different domains and use-cases. For example, \citet{hsu2007identification} analyzed algorithms for sequential pattern mining in the biomedical domain. 

In this work we use Markov chain models (see next section) as opposed to sequential pattern mining techniques for our experiments as they also allow us to directly gain insights into memory effects in our sequential data at interest. Furthermore, we can simply vary the length of patterns that we want to detect by changing the order of the Markov chain model.
}

\subsection{Markov chain models}
\label{sub:Markov chain models}

Previously, Markov chain models have been heavily applied for modeling Web navigation---some sample applications of Markov chains can be found in \cite{borges, deshpande, lempel, pirolli,sen, zukerman}.
Detailed specifications of the parameters used in a Markov chain---e.g., transition probabilities or also the specification of model orders---have previously been used to capture specific assumptions about the real human navigational behavior. One frequently used assumption is that human navigation on the Web is memoryless. This is further postulated in the \emph{Markovian assumption} which states that the next state in a system only depends on the current one and not on a sequence of preceding ones. This is, for example, also modeled in the Random Surfer model in Google's PageRank \cite{brin}.

Previously, researchers  have investigated whether human navigation really is memoryless in a series of studies (e.g., \cite{borges1999, pirolli}). However, they mostly have shown that the benefit of higher orders is not enough to compensate the extreme high number of parameters needed. Hence, the memoryless model seems to be a  plausible abstraction (see e.g., \cite{cadez, sarukkai, sen, zukerman}).
Recently, a study picked up on these investigations and again suggested that the Markovian assumption might be wrong for Web navigation patterns \cite{chierichetti}. 
Based on these controversies regarding memory effects in human navigation, \citet{singer_mc} presented a framework for determining the appropriate Markov chain order. Their studies on several navigational datasets  revealed that the memoryless model indeed seems to be a plausible abstraction.~ 
However, their work also highlighted that on a topical level (by looking at paths over topics instead of pages) clear memory effects can be observed. In this work, we adapt the corresponding framework in order to apply it to the process of collaborative ontology engineering.

Using Markov chains we want to learn more about the ongoing processes when collaboratively engineering an ontology, thus the work presented in this paper partly builds upon this and related lines of research and tries to expand them towards collaborative ontology authoring systems.

\section{Datasets}
\label{sec:datasets}
In this section, we present the main data studied in this paper. Mainly, we focus on the International Classification of Diseases (ICD-11) (Section~\ref{icd11}). For deriving the change-logs, we utilize the Change and Annotation Ontology (ChAO) (Section~\ref{sub:chao}).

\subsection{International Classification of Diseases, 11$^{\textnormal{th}}$ Revision}
\label{icd11}

ICD-11\footnote{http://www.who.int/classifications/icd/ICDRevision/}, developed and maintained by the World Health Organization, is the international standard for diagnostic classification that is used to encode information relevant to epidemiology, health management, and clinical use. 
Health officials use ICD in all United Nations member countries to compile basic health statistics, to monitor health-related spending, and to inform policy makers. As a result, ICD is an essential resource for health care all over the world. 

The development of \mbox{ICD-11} represents a major change in the revision process. Previous versions were developed by relatively small groups of experts in face-to-face meetings. \mbox{ICD-11} is being developed via a web-based process with many experts contributing to, improving, and reviewing the content online. It is also the first version to use OWL as its representation format.

We choose ICD-11 as an example ontology to demonstrate the effectiveness of the Markov chain methodology as the ontology satisfies several critical requirements for the applicability of our method: (i) at least two users have contributed to the project, and (ii) a structured log of changes (see section~\ref{sub:chao}) \marked{without ambiguous references to the elements in the ontology} is available. 
These characteristics can be seen as the minimum requirements to allow for an application of Markov chains onto collaborative ontology-engineering projects. For a list of characteristics for ICD-11 see Table~\ref{tab:dataset details}.

\begin{table}[b!]
\center
\scriptsize
\caption{\textbf{Characteristics of the International Classification of Diseases 11$^{\textnormal{th}}$ revision (ICD-11)} that we used for the demonstration to extract sequential patterns in collaborative ontology-engineering projects. The number of users corresponds to the number of users that have contributed at least $1$ change to ICD-11.}
\begin{tabular}{| l | c |}
\cline{2-2}
\multicolumn{1}{c|}{} & ICD-11 \\\hline
 concepts & 48,771\\
 changes & 439,229\\
 users & 108\\
\hline\hline
development tools & iCAT\\
\hline\hline
first change & 18.11.2009\\
last change & 29.08.2013\\
log duration (ca.)& 4 years\\
\hline
\end{tabular}
\label{tab:dataset details}
\end{table}

\begin{figure*}[th]
\centering
{\includegraphics[width=\linewidth]{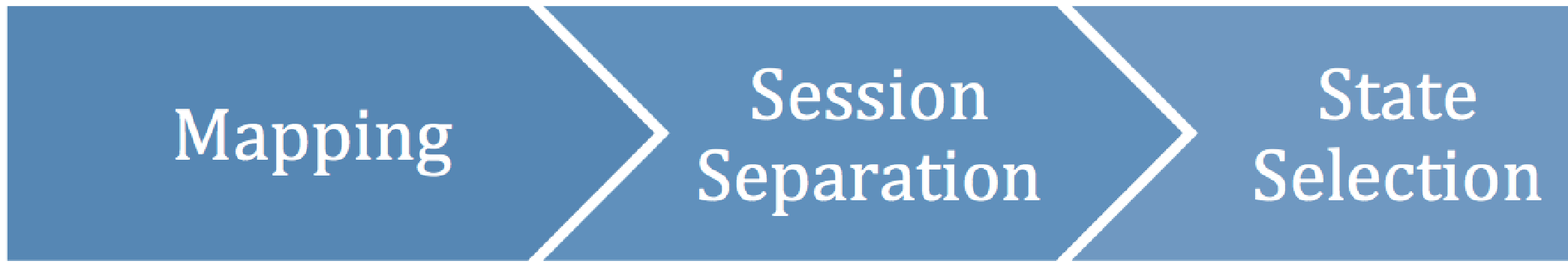}}
\caption{\textbf{The Analysis Process.} 
This figure depicts the different steps of the process that have to be performed to determine and evaluate the best fitting order of a Markov chain for a given dataset. The first two steps of the process involve a \emph{Mapping} (section~\ref{sub:mapping}) of the change-log data onto the underlying ontology and \emph{Session Separation} (section~\ref{sub:model separation}) tasks. The \emph{State Selection} step (section~\ref{sub:state selection}) is split into two separate tasks. First, questions have to be formulated that are to be investigated relying on the presented Markov chain analysis. 
Second, features of changes, which correspond to the previously formulated questions, have then to be identified and selected. In the \emph{Path Extraction} (section~\ref{sub:path extraction}) step, all of the previously identified features of changes have to be extracted and chronologically sorted.
Once the paths are extracted, they can be used as input for the \emph{Model Fitting} (section~\ref{sub:model fitting}), where the transition probabilities for the Markov chains are calculated. In the \emph{Model Selection} step (section~\ref{sub:model selection}), we determine the best fitting order of a Markov chain according to over- and under-fitting of the underlying data. The last step of the process, \emph{Interpretation} (section~\ref{sub:interpretation}), is used to combine the results of the different approaches of the \emph{Model Selection} to determine the best-fitting Markov chain order for the underlying data.}
\label{fig:process}
\end{figure*}

\subsection{The Change and Annotation Ontology (ChAO)}
\label{sub:chao}

The ontology that we use for the demonstration of the Markov chain-based sequential usage pattern analysis, \marked{the International Classification of Diseases in its $11^{\textnormal{th}}$} revision, is created using a custom tailored versions of \wpro called \emph{iCAT}. The tool provides a web-based interface as well as change-logs, which can be directly mapped onto the ontology that is to be created. The mapping of the change-log entries and the ontology depends on the availability of unique IDs for entities, such as users and concepts. These unique IDs are internally (unambiguously) mapped to the IDs (or URIs) of the corresponding elements of the ontology, allowing us to track, extract and analyze changes of concepts even if, for example, their title and all of their attributes are changed \marked{or their values are ambiguous}. This means that for every entry in the change log we have unique IDs that can be used to retrieve all involved entities. In traditional change-logs, which are usually separated from the productive environment, one minimalistic change could, for example, solely consist of one string, such as ``\emph{The title of concept 02 II Neoplasms was changed from Neoplasm to Neoplasms}''. The change logs provide a direct mapping to the concept and user (among others) affected by the changes, avoiding ambiguity, even if multiple concepts exhibit the same property values \marked{(i.e., have the same title ``Neoplasms'')}.
Note that whenever we refer to the underlying ontology, we refer to \marked{ICD-11 and not \chao or the change-logs.}

\begin{figure}[t]
\centering
{\includegraphics[width=\linewidth]{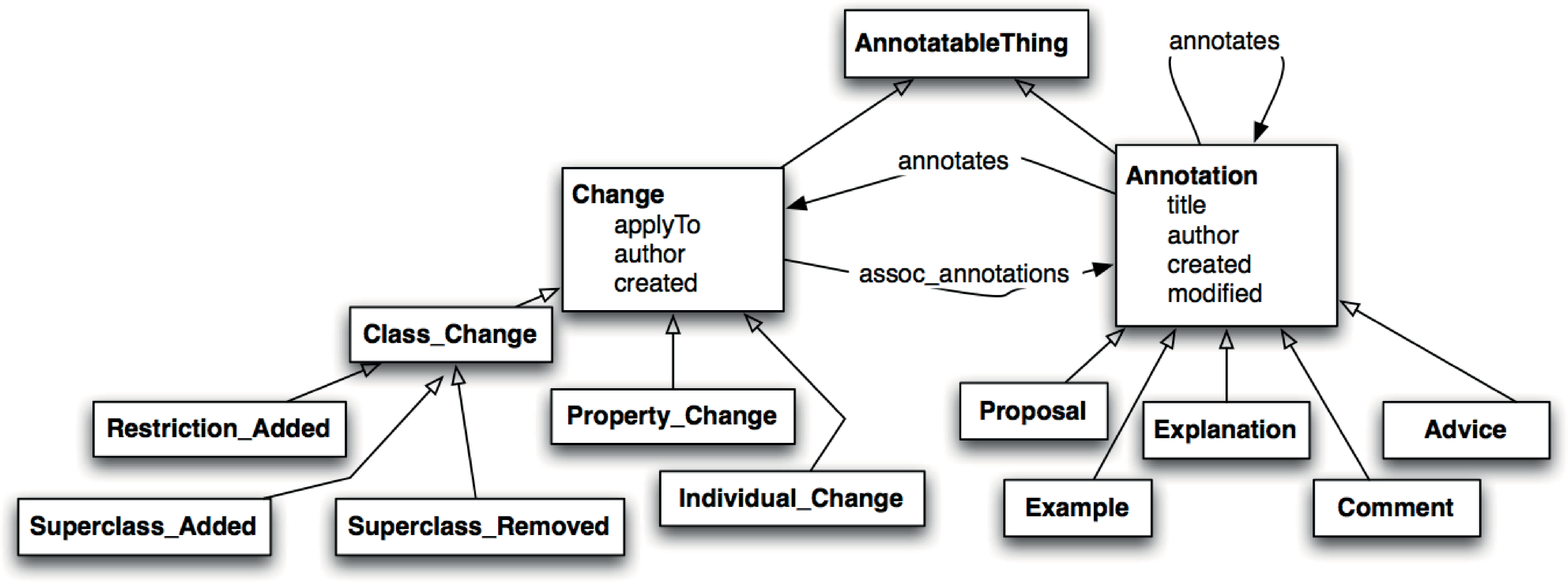}}
\caption{\marked{\textbf{The Change and Annotation Ontology.} The figure depicts a visual excerpt of the structure of the Change and Annotation Ontology (ChAO) used by \pro\ ~\cite{Noy.ISWC06}. Boxes represent classes and lines with arrows represent relationships (labeled) and subclasses.}}
\label{fig:chao}
\end{figure}

\pro and all of its derivatives use the \textbf{Change and Annotation Ontology (\chao)}~\cite{Noy.ISWC06} to represent these changes. \marked{In contrast to traditional change-logs, \chao itself represents a \emph{structured} log of changes that allows for explicitly (semantically rich) defined classes, properties and relationships. This means that change types are represented as ontology classes in \chao and changes in the domain ontology (e.g., ICD-11) are instances of these classes (Figure~\ref{fig:chao}).  Similarly, notes that users attach to concepts or threaded user discussions (represented as \emph{Annotations} in Figure~\ref{fig:chao}) are also stored in \chao. Further, \chao contains unique and unambiguous references to all entities in the ontology, for which \chao is storing the changes and annotations.}

\chao records two types of changes, so-called ``Atomic'' and ``Composite'' changes. 
``Atomic'' changes represent one single action within the ontology and they consist of several different types of changes such as \textit{Superclass Added}, \textit{Subclass Added} or \textit{Property Value Changed}. ``Composite'' changes combine several atomic changes into one change action that usually corresponds to a single action by a user. For example, moving a concept inside the ontology is represented by one composite change that consists of---at least---four ``atomic'' changes for removing and adding parent and child relations for all involved concepts. Every change and annotation provides information about the user who performed it, the involved concept or concepts, a time stamp and a short description of the changed or annotated concepts/properties. Whenever we talk about changes we refer to the $439,229$ changes stored in the \chao (see Table~\ref{tab:dataset details}), which are always actual changes to the ontology \marked{(e.g., changes performed on ICD-11; opposed to proposed changes)}.

\section{The Analysis Process}
\label{the process}

Figure~\ref{fig:process} depicts an abstraction of all the steps necessary to better understand the process
of how users sequentially edit properties of concepts in collaborative ontology-engineering projects. The first two steps of the analysis process, \emph{Mapping} (section~\ref{sub:mapping}) and \emph{Session Separation} (section~\ref{sub:model separation}), involve a mapping of the structured logs of changes onto the ontology 
as well as session separation tasks to prepare the data. In the \emph{State Selection} step (section~\ref{sub:state selection}) research questions are formulated allowing for the corresponding features of changes to be identified and selected. In the \emph{Path Extraction} (section~\ref{sub:path extraction}) step, all of the previously identified features have to be extracted and chronologically sorted
as they are needed as input for the Markov chain analysis. 

For the \emph{Path Extraction} step, we already have to know which questions we want to have answers for, as this determines the features of the changes that we are going to extract.
Once the change data is mapped, extracted and converted into the required format, we can start the \emph{Model Fitting} (section~\ref{sub:model fitting}). In this step, we use the extracted and preprocessed data to calculate the transition probabilities for the different orders of the Markov chain models. To determine which Markov chain order provides the best trade-off between model complexity and predictive performance we conduct several \emph{Model Selection} tasks (section~\ref{sub:model selection}).
In the last step of the process, \emph{Interpretation} (section~\ref{sub:interpretation}), we combine the gathered information of the model selection tasks and provide insights on choosing the Markov chain order that statistically significantly best models the sequential data.

\subsection{Step 1: Mapping}
\label{sub:mapping}

\marked{Given the structured nature of \chao, it already provides the necessary internal IDs to map the referenced entities, which are involved in the corresponding stored change-actions, to the corresponding concepts, properties and users of the actual ontology (for more details see section~\ref{sub:chao}). For example, if a specific property of a specific concept was changed, \chao would provide us with the necessary IDs to unambiguously identify the changed concept and property.} Hence, the mapping process for ICD-11 consists of simple id look-ups and joins between entries of \chao and the actual ontology. For other datasets, individual mapping strategies have to be developed or derived, which allow for an unambiguous identification of all involved entities, such as users, concepts or properties.

\subsection{Step 2: Session Separation} 
\label{sub:model separation}

Ontologies of the size of ICD-11 cannot be developed in one single day, hence we decided to introduce what we call \emph{artificial session breaks} to be able to gather more detailed information of the ongoing processes. 
As neither iCAT nor \chao provide information about user sessions, we manually added these artificial session breaks, which allow us to identify (or at least approximate) concepts and properties that users will work on, after or shortly before they take a break from editing the ontology.
These session break states are named \emph{BREAK} throughout all of our analyses and are specifically used to uncover the states before and after a break occurs in the change-logs for all analyses that investigate user-based activities (opposed to concept-based activities, which are only analyzed in section~\ref{sub:user-interface tab paths}).

\begin{figure}[!t]
\centering
\includegraphics[width=\plotwidth\linewidth]{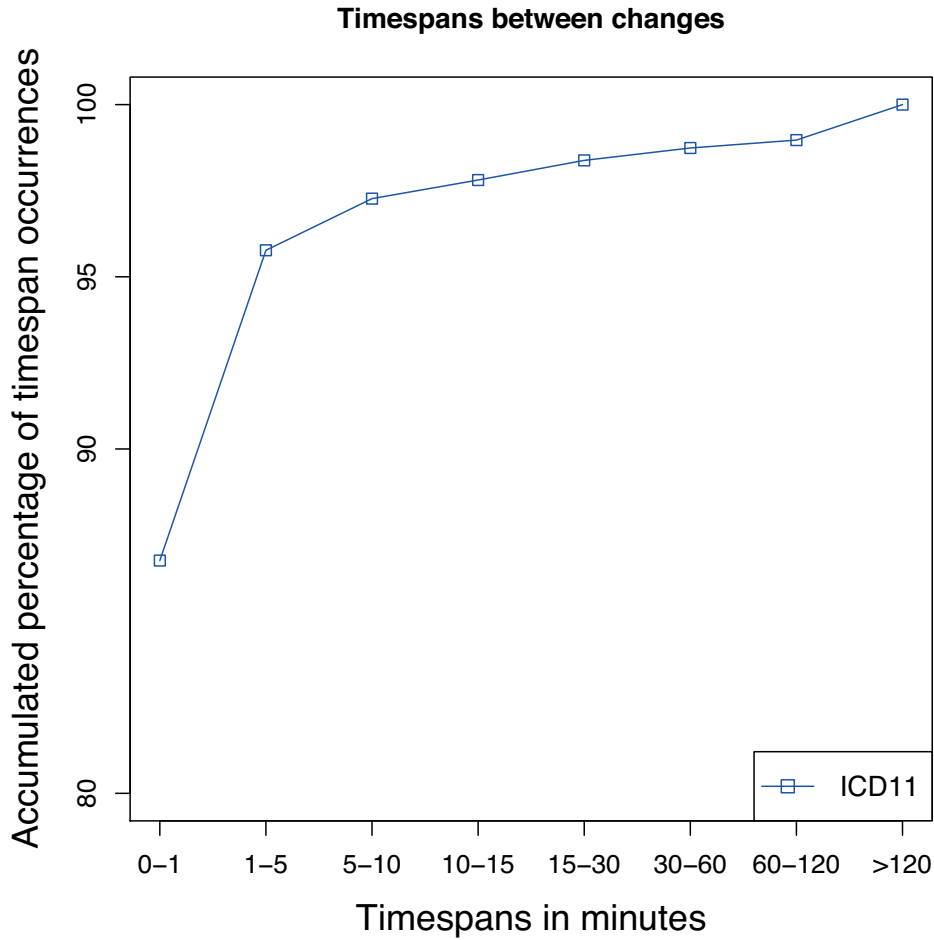}
\caption{\textbf{Occurrence of different timespans.} This plot depicts the percentage of all changes that have been performed within a specific timespan for ICD-11. The $x$-axis lists the timespans in minutes and the $y$-axis lists the accumulated percentage of all timespans between two consecutively conducted changes for every user. To avoid the introduction of too many \emph{artificial session breaks}, we decided to insert breaks for timespans between changes that are greater to the timespan so that $>95\%$ of all changes do not introduce new sessions. In the case of ICD-11, this timespan is the $1-5$ minutes one, meaning that \emph{BREAK}s have been introduced if the two changes in question are apart longer than $5$ minutes.}
\label{fig:timespans}
\end{figure}

Figure~\ref{fig:timespans} depicts the total amount of timespans between the changes of each user for ICD-11. The $y$-axis depicts the percentage of all changes performed within the corresponding timespan on the $x$-axis. The $x$-axis depicts the different timespan intervals in minutes. The majority ($>95\%$) of all changes in ICD-11 are performed within $5$ minutes. Thus, if two changes of the same user are apart longer than $5$ minutes, we have introduced an \emph{artificial session break} represented as a \emph{BREAK} state in all the conducted user-based analyses. 

\subsection{Step 3: State Selection}
\label{sub:state selection}
To be able to select the states for the Markov chain analysis we have to first define what kind of questions we seek answers for and then identify and extract the corresponding states. For example, if we are interested to know what kind of change a user is most likely to conduct next, the set of states to be extracted are all the different types of changes in the system. If we are interested in the relative movement of users, allowing us to predict if a user will move closer, further away or stay at the same distance to the root node, we have to extract the depth-levels of the changed concepts and compare the previous level with the current level to extract relative movement states (i.e., \emph{UP}, \emph{DOWN} and \emph{SAME}; for more info see section~\ref{sub:edit strategy paths}).

It is important to understand that, using Markov chains, we are mainly interested in predicting which state to occur next for a given user or a given concept. Note that if we do not have enough information to extract a chronologically ordered sequence of states, Markov chains cannot be used.

\subsection{Step 4: Path Extraction} 
\label{sub:path extraction}

To be able to analyze sequential usage patterns, we first have to extract sequential paths from the preprocessed structured logs of changes, which we can then use as input data for the Markov chains. 

\begin{figure}[!t]
\centering
{\includegraphics[width=0.7\linewidth]{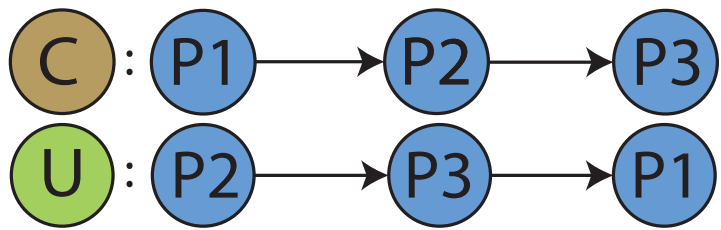}}
\caption{\textbf{Sequential Paths Sample.} The top row of the figure depicts an exemplary \textbf{concept-based} sequential property path ($P1$ to $P3$) for concept $C$. This means that for concept $C$ the property $P1$ was changed first, then property $P2$ and most recently changed was property $P3$. The bottom row of the figure depicts the sequential property path ($P1$ to $P3$) for a user $U$ (\textbf{user-based}). Analogously, user $U$ has first changed $P2$, continued to change property $P3$ and most recently changed $P1$.}
\label{fig:paths}
\end{figure}

A path represents a chronologically ordered list of changes or features that can be associated with that change, which are performed either by a user or are performed on a concept (Figure~\ref{fig:paths}). For example, when predicting the property that a user is most likely to work on next, we extract a chronologically ordered list of all changed properties for all users. We then store these lists in a file, where each user is represented by one line and the content of each line is the chronologically ordered list of changed properties of that user.

If we want to predict which property is most likely to be changed next for a given concept, we have to collect a chronologically ordered list of changed properties for each concept. Again, each line of the resulting file represents a concept while the content of each line is the chronologically ordered list of changed properties for that concept, \emph{not} including \emph{artificial session breaks} as this analysis is now concept-based.

For some of our analyses, we merged multiple consecutive changes of the same user on the same concept into two consecutive changes, resulting in \emph{one} \emph{self-loop}. For example, if one user would change the same property (e.g., title) on the same concept $5$ times, we would merge these $5$ changes of the same property into two changes, resulting in one \emph{self-loop} in the extracted path from title to title, opposed to four transitions from title to title. We performed this process of merging multiple consecutive changes into one single \emph{self-loop} to minimize the detection of higher order Markov chains that are biased towards transitions between the same states from the same concepts. 
This is particularly useful as there is no, or only minimal, actionable information when predicting that a user is going to perform the same change on the same concept again. \marked{If an ontology would provide multilingual properties and we are specifically interested in potential change-sequence patterns between these multilingual property values, we would have to create additional states accordingly (e.g., $property\_eng$, $property\_ger$, etc.)}

\subsection{Step 5: Model Fitting}
\label{sub:model fitting}

Markov chain models are well-known tools, among others, for modeling navigation on the web. 
We resort to and recapitulate the established methods first described by \citet{singer_mc}. 

In general, a Markov chain consists of a finite \emph{state-space} and the corresponding \emph{transition probabilities} between these states. 
For our analysis, we will make use of the transition probabilities to identify likely transitions for a variety of different states. To be able to do so, it is important to understand the nature of Markov chains. Formally, a finite and discrete (in time and space) Markov chain can be
seen as a stochastic process that contains a sequence of random variables--$X_1, X_2, \ldots, X_n$. One of the most well-known assumptions about
Markov chains is the so-called \emph{Markovian property} that postulates that
the next state of a sequence  depends only on the current state and not on a sequence of
preceding ones. Such a first-order (also called memoryless) Markov chain holds if:

\begin{eqnarray}
\nonumber P(X_{n+1} = x_{n+1} | X_1 = x_1, X_2 = x_2, ..., X_n = x_n) & = \\
 P(X_{n+1} = x_{n+1} | X_n = x_n)
\end{eqnarray}

We assume \emph{time-homogeneity} which means that the probability of a transition is independent of $n$.  For all our Markov chains and for
simplification we will refer to data (i.e., sequential paths) on which we fit a Markov chain model as a sequence $D=(x_1, x_2, ..., x_n)$ with
states from a finite set $S$. Hence, we can rewrite the Markovian property as:

\begin{equation}
 p(x_{n+1}|x_1, x_2, ..., x_n)=p(x_{n+1}|x_n)
\end{equation}

Furthermore, as we are also  interested in higher order Markov chains (i.e., the next state not only depends on the current one but on a series of preceding ones), we can state that in a $k$-th
order Markov chain the next state depends on $k$ previous ones. This leads to the following, more general equation: 

\begin{equation}
 p(x_{n+1}|x_1, x_2, ..., x_n)=p(x_{n+1}|x_n, x_{n-1}, ..., x_{n-k+1})
\end{equation}

Note that 
we can easily convert higher order Markov chains to
first-order Markov chains by modeling all possible sequences of length $k$ as
states and adjusting the probabilities accordingly. Hence, we can focus on
defining the methods for first-order chains solely, as this applies for higher
ones as well.

A Markov chain model is usually represented via a stochastic transition matrix
$P$ with elements $p_{ij}=p(x_j|x_i)$ where it holds that for all $i$:

\begin{equation}
 \sum_j p_{ij} = 1
\end{equation}

For easier understanding, one could think of a first-order Markov chain model as a matrix, where each column and row correspond to a state of the \emph{state-space} and the elements within the matrix represent the transition probabilities to and from each state towards the corresponding other states. For higher order Markov chain models, the states would include the combinations of all states, which is drastically increasing the state-space and thus, the complexity of the Markov chain. 

Furthermore, we also allow $k$ to be zero, resulting in a so-called \emph{zero-order} Markov chain model. This can be seen as a lower baseline and corresponds to a \emph{weighted random selection} \cite{singer_mc} -- i.e., the probabilities are defined by the number of occurrences of states.

\textbf{Maximum Likelihood Estimation (MLE):} To be able to determine the transition probabilities $p_{ij}$ between the states $x_i$ and $x_j$, we apply Equation~\ref{eq:trans-prob}, where $n_{ij}$ corresponds to the total number of transitions between states $x_i$ and $x_j$:

\begin{equation}
 p_{ij} = \frac{n_{ij}}{\sum_j n_{ij}}
  \label{eq:trans-prob}
\end{equation}

Hence, the maximum likelihood estimate (MLE) for the transition probability
$p_{ij}$ simply is the number of times we observe a transition between state
$x_i$ to state $x_j$ in our data $D$ divided by the total number of outgoing
transitions from state $x_i$ to any other state.

\subsection{Step 6: Model Selection}
\label{sub:model selection}

\marked{
As our goal is to determine the most appropriate Markov chain order, we need to
establish some methods for choosing the right one. Basically, we always want to compare a null model with an alternative model. To give an example, in our case the null-model could refer to a first-order Markov chain model while the alternative-model could refer to a second-order Markov chain model. Simply comparing likelihoods of two alternative models with each other is not enough though. Higher-order Markov chain models are always better fits to the data compared to lower-order ones by definition. This is reasoned by the higher complexity (higher number of parameters) of such higher-order Markov chain models. Thus, we need to balance the goodness of fit with the corresponding complexity when we want to compare models with each other.

To do so, we first focus on the \emph{Akaike information criterion (AIC)} and \emph{Bayesian information criterion (BIC)} to compare varying order Markov chain models with each other. In the following, we describe both methods, but we want to guide the reader to the work by \citet{singer_mc} for a more thorough description.
}
\smallskip

\noindent \textbf{Likelihood Ratio:} 
To be able to calculate AIC and BIC, we have to calculate the likelihood ratio, which 
simply is the ratio of the maximum likelihoods of the alternative and the null model.
The ratio gives us an indicator quantifying how much more likely the observed data is with the alternative model compared to the null model. As a result, we always compare lower
order models with higher order models. 
In order to avoid underflow, we calculate the log likelihood ratio.
We follow the notation by Tong
\cite{tong1975} who defines the log likelihood ratio as ${_k}\eta{_m}$:
\begin{equation}
 {_k}\eta{_m} = -2(\mathcal{L(P(D|}{\theta}_k)) -
 \mathcal{L(P(D|\theta}_m)))
 \label{eq:lr}
\end{equation}

$\mathcal{L(P(D|}{\theta}_k)$ represents the MLE for the null-model, while $\mathcal{L(P(D|}{\theta}_m)$ represents the MLE for the alternative model. 
\marked{
Note that simply using this likelihood ratio as a proper indicator for choosing between two models is not enough due to the reasons outlined above. Hence, we resort to the AIC and BIC methods which we outline next.}

\smallskip
\noindent \textbf{Akaike information criterion (AIC):} 
This information criterion can help us to determine the optimal
model from a class of competing models -- i.e., the appropriate Markov chain
order. The final method is
based on the minimization of the AIC -- minimum AIC estimate also called
MAICE -- \cite{gates1976} and has been first used for
Markov chains by Tong \cite{tong1975}. We define the AIC based on the work by Tong \cite{tong1975}:
\begin{equation}
 AIC(k) = {_k}\eta{_m} - 2(|S|^m-|S|^k)(|S|-1)
 \label{eq:aic}
\end{equation}

Basically, AIC subtracts the degrees of freedom from the likelihood ratio---thus, it penalizes models by their complexity. In our analysis, the degrees of freedom ($2(|S|^m-|S|^k)(|S|-1)$) represent two times the difference between the number of parameters for the null-model (order $k$) and the alternative model (order $m$). The basic idea is to choose $m$ as the maximum order we want to study and compare it with lower order models until the optimal Markov chain order is found. The most appropriate one is the one that exhibits the lowest AIC score.

\smallskip
\noindent \textbf{Bayesian Information Criterion (BIC):} 
This information criterion is very similar to the AIC except for the difference in penalization, as it increases negative weight on higher order models even more \cite{katz}:
\begin{equation}
 BIC(k) = {_k}\eta{_m} - (|S|^m-|S|^k)(|S|-1)ln(n)
 \label{eq:bic}
\end{equation} 

We proceed similar as for AIC and choose $m$ reasonably high. The specific penalty function  is the degree of freedoms multiplied with the natural logarithm of the number of observations $n$ \cite{katz}, where an observation is always represented as a state in the change-logs. 

\smallskip
\textbf{Prediction Task:} In addition to our information-theoretic methods for determining the appropriate Markov chain order, we use a cross validation prediction for this task. This prediction task is conducted to actually measure which model order is best suited for predicting the next state, with the available change-logs as input. The main idea behind this approach is to calculate the parameters on a training set and to validate the model on an independent test set. We apply Laplace smoothing in order to be able to predict states that are  present only in the test set and not in the training set. To reduce variance, we perform a stratified 7-fold cross validation. In this case, we stratify the folds in order to keep the number of visited states in each fold equal. 

The validation  is based on the task of predicting the next step in a path of the test set. This validation also enables us to get detailed insights into the prediction possibilities of distinct Markov chain order models. Simply, one could predict the next state by taking the state with the highest probability in the transition matrix $P$. 
In the following, we describe the process of calculating the prediction accuracy.

\marked{
First, we start by calculating the prediction accuracy for each fold separately. For doing so, we determine the average rank  of each observation in a test set given the transition matrix as learned from the training data. In detail, given a current state $x_n$ (or series of preceding states for higher order models), we look up the rank of the next state $x_{n+1}$ in the sorted list of transition probabilities.
Next, we average over the rank of all observations in the test set. We follow the notation of \citet{singer_mc} and define the average rank $\overline{r(D_f)}$ of a fold
$D_f$ for some model $M_k$ the following way\footnote{alternatively, one could also use measures like perplexity}:
\begin{equation}
 \overline{r(D_f)} = \frac{\sum_i \sum_j n_{ij} r_{ij}}{\sum_i \sum_j n_{ij}},
\end{equation}

where $n_{ij}$
is the number of transition from state $x_i$ to state $x_j$ in $D_f$ and $r_{ij}$ denotes the rank of $x_j$ in the $i$-th row of $P$.
As frequently ties occur in these rankings, we assign the maximum rank to such ties (i.e., modified competition ranking). This method also includes a natural Occam's razor (penalty) for higher order models.
After we have calculated the prediction accuracy of all folds, we average them and suggest the model with the lowest average rank.
}

In the last part of the \emph{Model Selection}, we have to manually assess and combine the different results from the information criteria, the significance tests and the prediction task (see section~\ref{sub:model selection}), to determine the Markov chain order, which provides the best trade-off between model complexity (and thus, also computation time) and predictive power. 
Depending on the size of the change-log and the number of states that we want to investigate and predict, the different information criteria yield different suggestions for the best fitting Markov chain order, avoiding over- and under-fitting. The significance tests provide information about the highest Markov chain order, that is still significantly different to the remaining Markov chain orders.

In general, BIC exhibits a tendency to suggest lower Markov chain orders than AIC, due to the heavier weighted bias on model complexity. In contrast, the prediction task usually suggests the usage of higher order Markov chains. However, on closer investigation, the absolute differences between the suggested orders of AIC and BIC versus the suggested order of the prediction task, most of the time, do not justify the drastically increased model complexity (and thus computation time) of higher order Markov chains.

\marked{Overall, all presented methods try to achieve the same goal, i.e., balancing the goodness of fit with the number of parameters of varying Markov chain orders. Higher order Markov chain models have much higher complexity and thus, are potentially prone to overfitting. AIC and BIC achieve this in a natural manner by having direct complexity balance terms in their equations. For cross-validation, we try to include a natural Occam’s razor by our corresponding choice of how to rank ties. Thus, we believe that contrasting all presented methods in this article provides really thorough insights into the appropriate Markov chain order given the data. 

However, as mentioned, the results of these methods (which frequently match anyhow) might be weighted differently according to the goal of the study. If the main goal is to study predictability, one might want to focus on cross-validation as it also directly provides a measure of how well we can predict with varying order models. However, the calculation of the cross-validation is quite expensive, which is why one want to focus on AIC and BIC. The focus of these two methods is to provide an answer to how well varying order models fit the data in relation to each other. As mentioned, complexity is incorporated; BIC has a higher penalty for complexity compared to AIC. According to \citet{singer_mc} , AIC might be best suited for prediction, while BIC might be better for explanation. This is also reasoned by the observation that AIC is asymptotically equivalent to cross validation if both use MLE. 
As a final note, we want to mention that BIC is asymptotically consistent. For further information of the advantages and disadvantages---as well as further methods for order estimation---please refer to the work by \citet{vrieze2012model} and \citet{singer_mc}.}

\textbf{Limitations:} 
\marked{
Note that the model-estimation methods described in this work balance the goodness of fit with the number of parameters needed for each Markov chain order model. This trade-off  is necessary, as specifically higher order models need an exponentially growing number of parameters which might not be offset by the statistically significant benefit against lower order models and is also reflected by the initial choice about the set of states to consider. 
Thus, the results are also influenced by the amount of finite data available which is a common problem of statistical methods that mostly rely on asymptotic approximations. Basically, the more data we observe, the more amenable we are towards more complex models---i.e., higher order Markov chain models.
Hence, if the underlying process actually accords to a higher order Markov chain process, we need a certain amount of data for a given complexity, to be able to properly detect this order. 
With insufficient data, lower orders might be identified as being appropriate as the goodness of fit cannot compensate the complexity. 
Hence, it is also necessary to have large change-logs available in order to have the opportunity to detect higher order Markov chain models. 

The required total number of available observations, that is the number of performed changes, for detecting potential higher orders is directly related to the number of unique states that are extracted. For example, if all changes are mapped on two unique states (e.g., \emph{structural changes} and \emph{property changes}), smaller change-logs might already yield satisfying results, whereas higher numbers of unique states might require exponentially larger change-logs for the detection of higher orders.

In this work (see Table~\ref{tab:dataset details}), we study a dataset with around $440,000$ changes and with a limited number of distinct states. Also, our results highlight several higher order models as being more plausible compared to lower order models. Thus, we can be confident that we have sufficient data to detect higher order Markov chain models as being appropriate, if they actually are. If a zero order Markov chain model would be suggested each time, we would need to rethink our data base.
}

\subsection{Step 7: Interpretation}
\label{sub:interpretation}

After determining the best fitting Markov chain order we can start interpreting the results. For example, when investigating the next, most likely change type to be performed by a user, we can look at the transition probabilities and given $n$ previous changes, where $n$ equals the order of the best fitting Markov chain model, infer a ranked list of most probable transitions.

\section{Demonstration \& Evaluation}
\label{evaluation}

In this section, we investigate the qualitative analysis that we can do using sequential pattern analyses. We present the types of questions that we can ask and provide the example analysis based on the change logs for the editing process of  ICD-11~(Table~\ref{tab:dataset details}). 

In section~\ref{sub:change type paths}, we investigate if and to what extent sequential patterns of performed change types can be detected.

To see where and how users contribute to the ontology and if they exhibit sequential patterns when doing so, we analyze edit strategy patterns, such as  \emph{bottom-up} or \emph{top-down} editing behavior (section~\ref{sub:edit strategy paths}).

In section~\ref{sub:user-interface tab paths}, we report on our investigation on whether users have to  switch frequently between  different sections of the user-interface while contributing to ICD-11 and how often (and in which order) do they use the  different sections of the user-interface in order to add information for a concept.

\textbf{Step 1, Mapping} and \textbf{Step 2, Model Separation} are the same for all types of analyses that we present in this section. We describe these steps in sections~\ref{sub:mapping} and \ref{sub:model separation}. In the remainder of this section we focus on the remaining steps, which differ from one type of paths to the other. 
\textbf{Step 7, Interpretation} is mainly focusing on the implications of the best fitting order of Markov chains, rather than an in-depth investigation of the transition probabilities. A detailed interpretation of the transition probabilities for the \emph{Change-Type Paths} analysis, can be found in section~\ref{discussion}.

\subsection{Change-Type Paths}
\label{sub:change type paths}

\begin{table}
  \centering
  \caption{\marked{\textbf{Change Types.} The table depicts all types of changes that are used in the \emph{Change-Type Analysis} in Section~\ref{sub:change type paths}} \blue{The change types MOVE and CREATE represent the corresponding changes performed on the classes. Note that classes in ICD-11 are not deleted, but are moved to specific areas in the ontology, hence the omission of the DELETE type.. As the different properties in ICD-11 have been determined very early on in the development process and additional properties are very rarely introduced---which can only be done by administrators---we have neglected these types of changes and concentrated our analysis on the different edit actions that can be performed on properties.}}
  \label{tab:ct_tab}
  \begin{tabular}{ l | l }
  \textbf{Change Type} & \textbf{Description} \\\hline
  MOVE & Changes that move a class. \\
  CREATE & Changes that create a new class. \\
  BOT & Changes that were automatically performed by bots.\\
  OTHER & Any change that does not fit any other change type.\\
  EDIT\_REPLACE & Changes that replace the value of a property.\\
  EDIT\_REMOVE & Changes that remove the value of a property.\\
  EDIT\_IMPORT & Changes that import the value of a property.\\
  EDIT\_ADD & Changes that add a value to a property.\\
  \hline
  \end{tabular}
\end{table}

\textbf{Step 3: State Selection.}
The analysis of change types provides information about the type of a change that a user will most likely conduct next. The information of what kind of change a user is most likely to perform next could be used by, for example, user-interface designers and ontology-engineering tool developers to automatically adapt the interface. Additionally, knowing if users primarily concentrate their efforts on the same change types or engage in multiple diverse actions while editing the content of the ontology can also be used by project administrators for curation purposes. Furthermore, when investigating the transition probabilities between the different change types, it is possible to identify certain pairs of changes that ``usually'' occur together, providing again information for automatic user-interface adaptions.

\textbf{Step 4: Path Extraction.} We aggregated the change types into more abstract change-classes to minimize the necessary state space for detecting Markov chains, which still provide useful information for curation and work-delegation purposes. \marked{Note that these change types only represent an abstracted fraction of all available change types in \chao.} In general, these change-type classes are \emph{CREATE} and \emph{MOVE}, which include all changes that have a corresponding effect on classes in the ontology. \blue{Note that classes in ICD-11 are not deleted, but are moved to specific areas in the ontology, hence the omission of the DELETE type.} Furthermore, we have created the classes \emph{EDIT\_ADD}, \emph{EDIT\_IMPORT}, \emph{EDIT\_REMOVE} and \emph{EDIT\_REPLACE}, which are used when values of properties are either added, imported, removed or replaced. There are two special cases for ICD-11, namely \emph{BOT} and \emph{OTHER}. The first change-type is used for automatically performed changes while the latter is used to mark changes that are not included in the other listed change-type classes, such as addition of direct types or adding and removing sub- and superclasses \marked{(see Table~\ref{tab:ct_tab} for a short description of all change types)}.

\blue{In general, all types of changes with the ``\emph{EDIT\_}'' prefix are changes performed on the properties of a class. As the different properties in ICD-11 have been determined very early on in the development process and new properties are rarely introduced---which can only be done by administrators---we have neglected these types of changes (i.e., are aggregated as part of \textit{OTHER}) and concentrated our analysis on the different edit actions that can be performed on existing properties.}

\blue{For creating the sequential paths, we first mapped all the changes of each user in our datasets to the different aggregated change-classes. In a second step we stored them as chronologically ordered lists for each user and each dataset individually. Multiple consecutive identical change types of the same user on the same concept were merged into one \emph{self-loop}.} 

\begin{figure}[!t]
\centering
\includegraphics[width=0.8\linewidth]{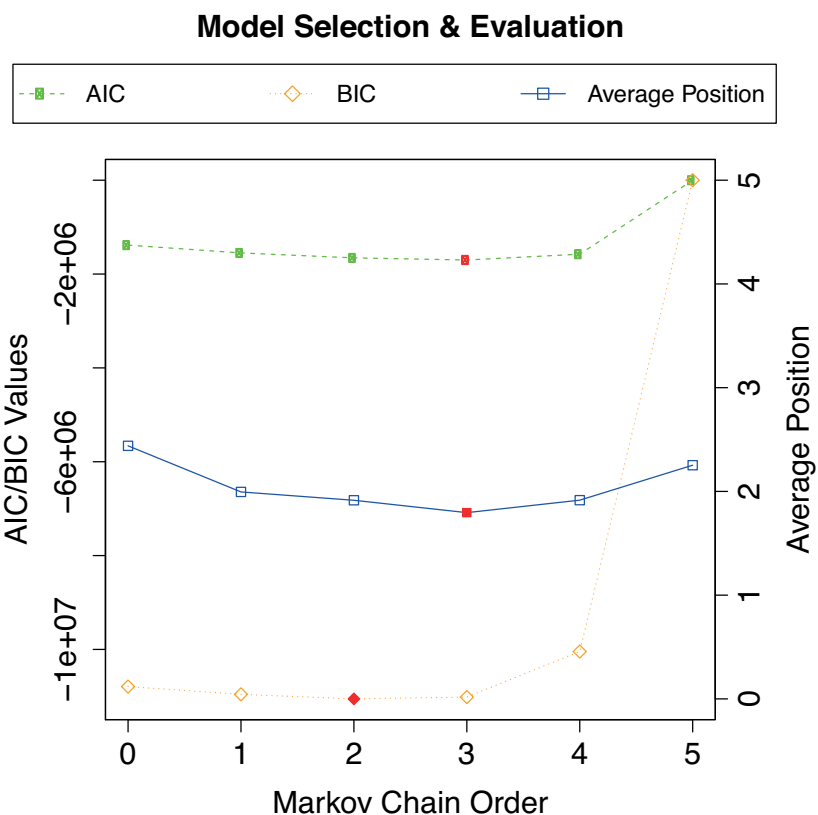}
\caption{\textbf{Change Type Paths Model Selection and Evaluation.} This plot depicts the results of the AIC and BIC model selection criteria as well as the stratified cross-fold evaluation for the \emph{Change Type Paths} analysis. 
The $x$-axis represents the different Markov chain orders. The left $y$-axis lists the AIC and BIC values of our model selection, while the right $y$-axis shows the average position values for the prediction task.
The filled elements represent the corresponding Markov chain models, which achieved the best (lowest) average position score in the prediction task or lowest AIC and BIC values for the model selection. The information criteria, AIC and BIC, suggest the usage of a third- and second-order Markov chain respectively. The prediction task performed best relying on the predictive information extracted from a third-order Markov chain.} 
\label{fig:ctuccv}
\end{figure}

\textbf{Step 5: Model Fitting \& Step 6: Model Selection.} We used the extracted paths  to calculate the transition probabilities between the different change-type classes in the \emph{Model Fitting} step. We then calculated AIC and BIC for the extracted Markov chain models of varying order (Figure~\ref{fig:ctuccv}) to identify the appropriate order that reflects to what extent contributors exhibit memory patterns when changing concepts.

 AIC and BIC suggest the usage of a third- and second-order Markov chain respectively. The likelihood ratio tests strengthen this observation as a second-order Markov chain for ICD-11 is significantly different from a first-order Markov chain, thus suggesting the selection of a second-order Markov chain model for predicting the next change type. 

To determine which order of a Markov chain contains the highest predictive power, we conducted a stratified cross-fold validation prediction task (see section~\ref{sub:model selection} for a detailed explanation). 
As depicted in Figure~\ref{fig:ctuccv}, the stratified cross-fold validation encourages the usage of a third-order Markov chain for \mbox{ICD-11}. 

The combined results of the model selection tasks indicate the best performance with the usage of a third-order Markov chain for ICD-11 for the task of predicting the change type a user is most likely to conduct next.

\textbf{Step 7: Interpretation.} 
A Markov chain of third order indicates that the last three change types a user has performed provide the best amount of information on the change type that is most likely to be performed next by that user. \marked{This information can (potentially) be used by programmers and designers of ontology development tools to automatically adjust parts of the interface according to the change-action a user is most likely to perform next. For example, if the next change will most likely involve deleting a concept the user-interface could already present and/or highlight specific parts that correspond to the anticipated action or display additional information, such as recently deleted concepts by the corresponding user. Note that these results are specific for ICD-11 and iCAT and might differ for other collaborative ontology-engineering projects.}

\subsection{Edit-Strategy Paths}
\label{sub:edit strategy paths}

\textbf{Step 3: State Selection.}
The analysis of \emph{Edit Strategy Paths} focuses on the investigation of relative movement along the ontological structure. Using the gathered data we can identify if users who are contributing to the ontology are more likely to follow a \emph{bottom-up} or \emph{top-down} editing strategy. For example, if users would create or edit an ontology in a \emph{bottom-up} manner, they would first model very specific concepts and continue to devote their work on more abstract concepts, while a \emph{top-down} approach would work the opposite way. Note  that this analysis can identify edit-strategy tendencies, however it could lead to wrong conclusions without manual verification of the change-logs. For example, if users generally tend to work on concepts in an alphabetical order, it is possible that this analysis could yield either, a \emph{bottom-up}, a \emph{top-down} or a non-apparent or random edit strategy, even though users do not purposely move along the semantic structure of the underlying ontology when contributing to the system. To make sure that our dataset does not exhibit such a behavior we have manually investigated the structured log of changes of ICD-11 to verify that the mentioned kind of contribution behavior is not present.

Furthermore, we were not able to recreate the exact class hierarchy of ICD-11 for every single change. This limitation is partly due to a lack of detail in the change-logs \marked{(e.g., some changes were conducted by the administrators of iCAT in the database, circumventing iCAT and \chao. Hence, no change-logs are available of these actions, preventing a complete reconstruction of the ontology at every point in time)}. Thus, we decided to use the ontology \emph{as is} at the latest point in time for our analysis. This basically means that if a class was changed by a user and afterwards moved, we would assume that this class has always been at the new location. To measure the extent of the potential bias, we counted all changes that were performed on a class before it was moved within in the ontology resulting in a total of 116,204 of 439,229 changes representing about 1/4 of all changes for ICD-11. 

In particular, this analysis allows us to predict if the concept a user is going to contribute to next is on the same, a lower (more abstract) or a higher (more specialized) hierarchy level of the ontology. Using the gathered information we can infer if users follow a \emph{top-down} or \emph{bottom-up} edit strategy while contributing to ICD-11.

\textbf{Step 4: Path Extraction.} The states in this analysis indicate if a user, when contributing to the ontology, moved either closer to (state \emph{UP}), further away (state \emph{DOWN}) or kept the same distance (state \emph{SAME}) from the root concept of the ontology. 

We gathered the sequences  for this analysis by calculating the shortest paths between all the concepts in the ontology and the root node, following \emph{isA}\footnote{\blue{For our analysis we only consider \emph{isA} relationships with regards to the \emph{rdfs:subClassOf} property. In particular, classes connected via (directed) \emph{isA} relationships specify that all the instances of one class (source) are also instances of the other class (target).}} relationships. For ICD-11 the root category is \emph{ICDCategory}, which is an equivalent of \emph{owl:Thing}. Again, we merged multiple \emph{self-loops}, represented by consecutive changes performed by the same user on the same concept, into one single transition. 
We have removed the data on users who contributed fewer than two changes from the analysis, as we require at least two changes to infer transitions between concepts.

A sample path is depicted in Figure~\ref{fig:icat}. When following the annotations $A-C$, which represent the changes performed by one user, we can extract the following path: \emph{DOWN, SAME, DOWN}. \blue{Note that for the creation of the first state we have to look at the first two classes that were changed by the corresponding user.}

\begin{figure}[!t]
\centering
\includegraphics[width=0.8\linewidth]{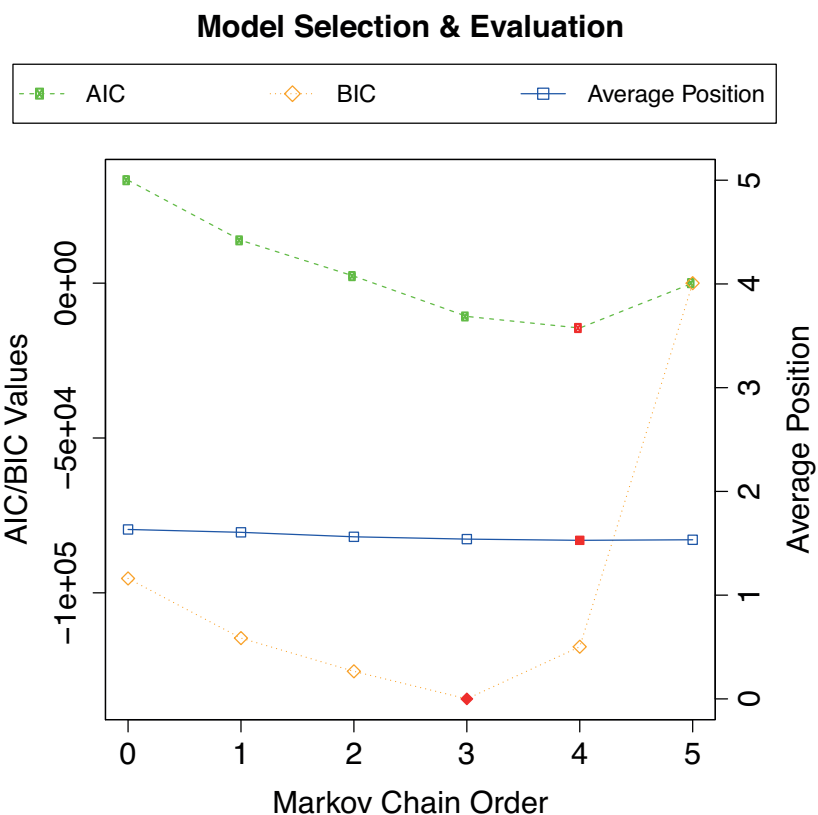}
\caption{\textbf{Edit Strategy Paths Model Selection and Evaluation.} This plot depicts the results of the AIC and BIC model selection criteria as well as the stratified cross-fold evaluation for the \emph{Edit-Strategy Paths} analysis. The $x$-axis represents the different Markov chain orders. The left $y$-axis lists the AIC and BIC values of our model selection, while the right $y$-axis shows the average position values for the prediction task. The filled elements represent the corresponding Markov chain models, which achieved the best (lowest) average position score in the prediction task or lowest AIC and BIC values for the model selection.
The information criteria, AIC and BIC, were able to detect a fourth- and third-order Markov chain respectively. 
The prediction task yielded the best results with a fourth-order Markov chain model.  
}
\label{fig:esccv}
\end{figure}

\textbf{Step 5: Model Fitting \& Step 6: Model Selection.} We used the extracted paths  to calculate the transition probabilities between the different change-type classes in the \emph{Model Fitting} step. We then calculated AIC and BIC for the extracted Markov chain models (Figure~\ref{fig:esccv}) to identify the appropriate Markov chain order when modeling edit-strategy patterns of contributors changing concepts. For ICD-11 both AIC and BIC suggest a fourth- and third-order Markov chain respectively.
Our likelihood ratio tests show that a third-order Markov chain for ICD-11 is still significantly different from a fifth-order Markov chain, indicating that either a third, fourth- or fifth-order Markov chain provides the best balance between model complexity and predictive power.

To determine the best-fitting Markov chain model orders to predict the next relative depth-level we conducted a stratified cross-fold validation prediction task (see Figure~\ref{fig:esccv}). 
The results of our prediction experiment suggest the usage of a fourth-order Markov chain for ICD-11. 

As the differences between the higher-order Markov chains and the third-order Markov chain are very small, yet different, we agree with BIC and the significance test on the usage of a third-order Markov model for predictive tasks, due to the high increase in complexity of the higher-order models.

\textbf{Step 7: Interpretation.} 
A Markov chain of first order indicates that the last relative depth-level of a change performed by a user provides better information on where the user is going to change a concept next (as relative depth-level) than randomly selecting either \emph{UP}, \emph{DOWN} or \emph{SAME}. \marked{After inspecting the resulting transition probabilities between the different states, we can conclude that users in ICD-11 exhibit a top-down edit strategy. Particularly, they are likelier to stay on the same or switch to a lower level of the ontology than they are, changing a class on a higher level of the ontology. In particular, this information could be exploited by project administrators to adjust milestones (i.e., first completing branches of the ontology, rather than adding properties to all concepts of the ontology). Note that these results are specific for ICD-11 and iCAT and might differ for other collaborative ontology-engineering projects.}

\subsection{User-Interface Sections Paths}
\label{sub:user-interface tab paths}

\textbf{Step 3: State Selection.}
The goal of this analysis is to investigate if we can map changes that occur in the ontology to actual areas and sections of the user-interface of iCAT, the collaborative ontology-engineering tool used to develop ICD-11. 
The user-interface of iCAT is divided into several sections, thematically grouping properties of concepts. For example, as depicted in Figure~\ref{fig:icat}, the user-interface section \emph{Title \& Definition} groups the properties \emph{ICD-10 Code}, \emph{Sorting label}, \emph{ICD Title}, \emph{Fully Specified Name} and \emph{Short Definition}. Other user-interface sections, grouping different properties, are for example, \emph{Classification Properties}, \emph{Terms} or \emph{Clinical Description}.
We investigate two different approaches: First, the user-based approach, where we analyze the sections of the user-interface used by contributors when editing the ontology. Second, the concept-based approach, where we investigate which sections of the user-interface are used when concepts are populated with data. If patterns can be detected, ontology-engineering tool developers can use this information to minimize the necessary effort for users to be able to contribute. It is important to note that not all properties and sections of iCAT are already actively used as ICD-11 is still under active development. \marked{Hence, the results of the presented analysis are limited by the properties and sections that are already available and actively used in iCAT. Rather than focusing on the results, this specific analysis was selected to demonstrate the feasibility and potential of the Markov chain analysis.}

\begin{figure*}[!ht]
\centering
\subfigure[User-based approach]{\label{fig:tuscv:a}\includegraphics[width=0.45\linewidth]{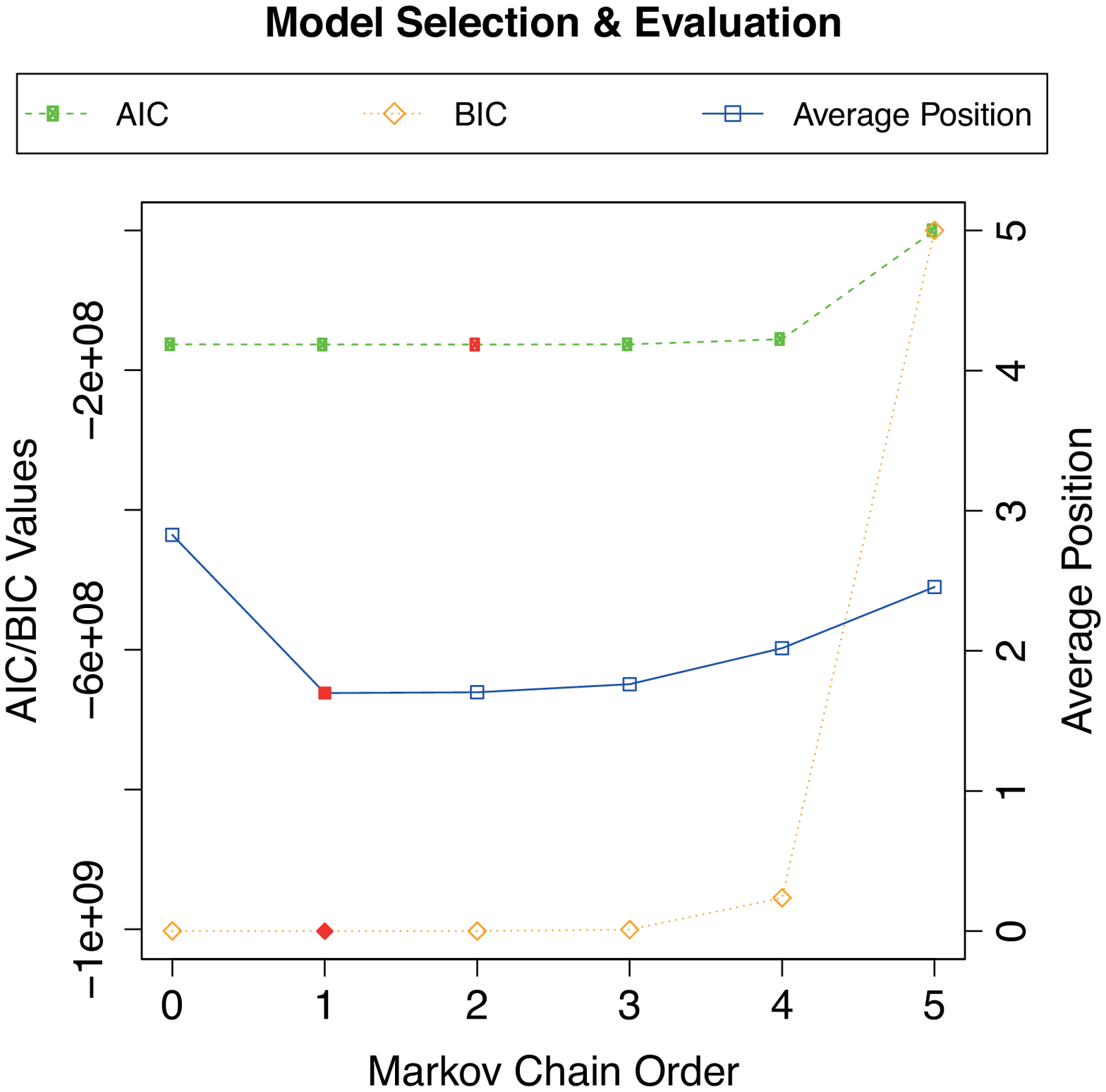}}
\subfigure[Concept-based approach]{\label{fig:tuscv:b}\includegraphics[width=0.45\linewidth]{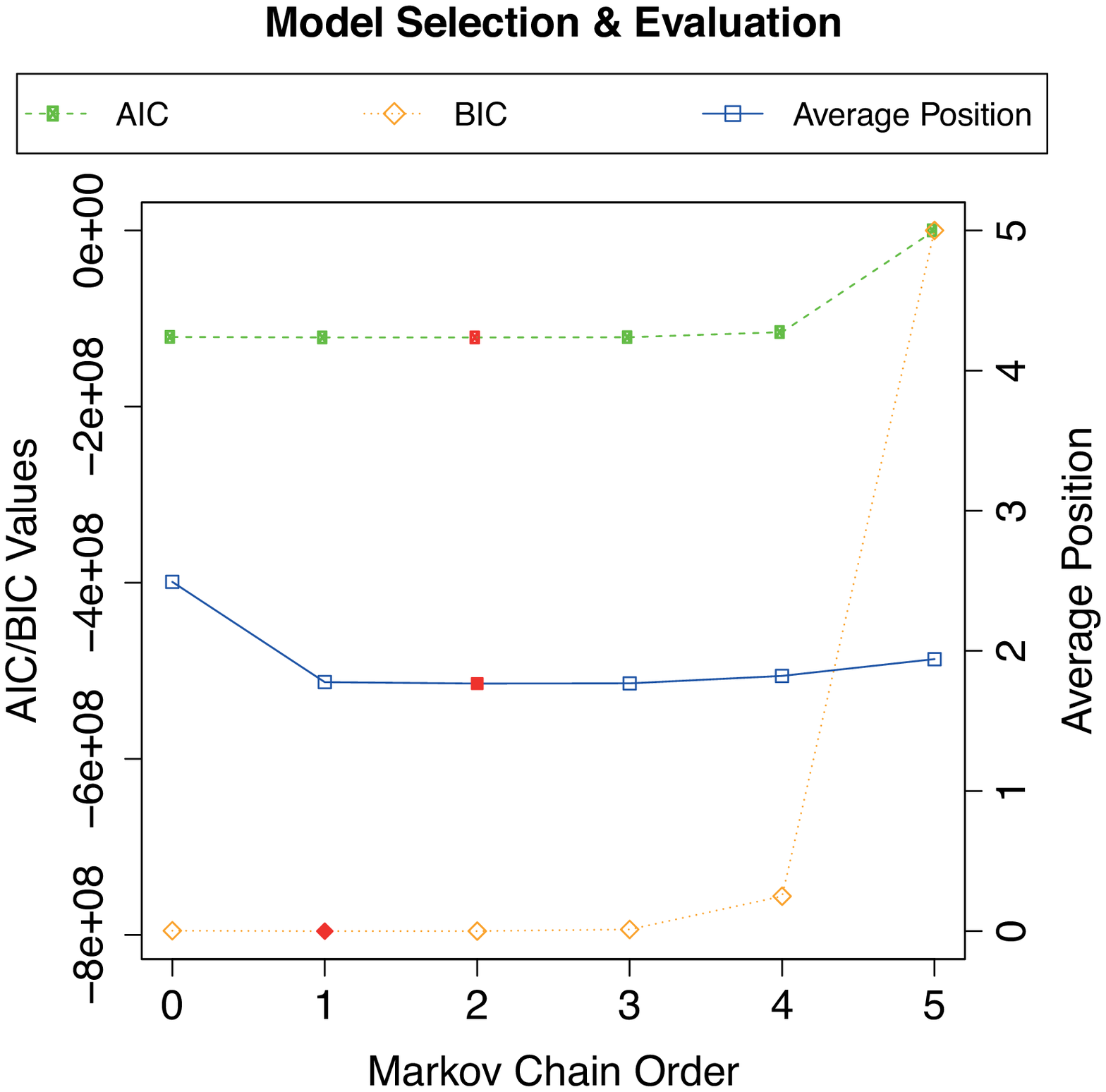}}
\caption{\textbf{User-Interface Sections Path Model Selection and Evaluation.} 
This plot depicts the results of the AIC and BIC model selection criteria as well as the stratified cross-fold evaluation prediction task for the user- and concept-based approaches of the \emph{User-Interface Sections Paths} analyses. The $x$-axes represent the different Markov chain orders. The left $y$-axes list the AIC and BIC values of our model selection, while the right $y$-axes show the average position values for the prediction task. The filled elements represent the corresponding Markov chain models, which achieved the best (lowest) average position score in the prediction task or best (lowest) AIC and BIC values for the model selection.
For both approaches, AIC and BIC suggest a second- and first-order Markov chain respectively, while the prediction task produced the best average position with a Markov chain of first-order for the user-based and second-order for the concept-based approach. 
}
\label{fig:tuscv}
\end{figure*}

\textbf{Step 4: Path Extraction.} The states for this analysis are represented by the different user-interface sections of iCAT. An excerpt of all different user-interface sections of iCAT can be seen in Figure~\ref{fig:icat}. 

To be able to analyze sequential patterns of different user-interface sections we extracted the chronologically ordered list of changed properties for (i) each user and (ii) each concept. We then continued by mapping the extracted properties to sections in the user-interface of iCAT. Whenever a change did not affect a property (e.g., because the change-action dealt with moving or creating a concept) and thus did not affect a user-interface section, the \emph{no property} state was used.
Analogously to the previous analyses, we merged consecutive changes of the same user on the same concept on the same property  into one \emph{self-loop} for the user-based analysis. For the concept-based analysis consecutive changes on the same concept and property have been merged into one \emph{self-loop}.

A sample path is depicted in Figure~\ref{fig:icat}. When following the annotations $I-III$, which represent consecutive changes performed by one user, using the highlighted sections of the user-interface, the following path can be extracted: \emph{Title \& Definition, Terms, Causal Properties}.

\textbf{Step 5: Model Fitting \& Step 6: Model Selection.} We calculated AIC and BIC for the extracted Markov chain models (see Figures~\ref{fig:tuscv:a} and \ref{fig:tuscv:b}) to determine the appropriate Markov chain order when modeling how users switch between sections of the interface when contributing to the ontology. For both approaches AIC and BIC suggest a second- and first-order Markov chain respectively.
The conducted significant tests show that a second-order Markov chain for both approaches is significantly different from a first-order Markov chain, indicating that either a second-order or first-order Markov chain provide the best balance between model complexity and predictive power.

To determine the predictive power of the investigated Markov chain models of varying orders for predicting the section most probably used to edit a property next, a stratified cross-fold validation prediction task (see Figure~\ref{fig:tuscv}) was conducted.
For the user-based approach a first-order and the concept-based approach a second-order Markov chain yielded the best predictions. 

Due to the fact that the determined second-order Markov chain performed nearly as well as a first-order Markov chain, it is best to use a first-order Markov chain to predict the next user-interface section, that a user is going to use, as it provides the best balance between model complexity (and thus computation time) and predictive power. 

\textbf{Step 7: Interpretation.} 
A first-order Markov chain indicates that the last user-interface section, used to conduct a change by a specific user, contains information about the user-interface section that this specific user is most likely to use for the next change. \marked{If we would observe high transition probabilities between a fraction and frequently used sections of the user-interface, this could indicate that users have to visit many different sections while following their normal work-flow. If our inherent goal was to increase activity and contributions, a first potential approach could involve the restructuring of the user-interface to better accommodate this inherent edit-process by reducing or even minimizing the required clicks (and hence time) to contribute. Note that the proposed applications and implications of our analyses are of theoretic nature, to highlight the potential of the Markov chain analysis process. For future work we plan on further analyzing, validating and evaluating the recommendations and predictions generated via our Markov chain analysis in live-lab studies for multiple different ontology-development tools.}

\begin{table*}[!hb]
\center
\scriptsize
\caption{\textbf{This Table depicts a summary of all gathered results for ICD-11 and the performed analyses} of section~\ref{the process}. The numbers in this table represent the calculated and suggested Markov chain orders from our model selection (AIC and BIC), significance tests (Significant Diff.) and evaluation tasks (Prediction Task). \emph{Best Balance} indicates the manually selected best-fitting order of a Markov chain, which represents the best trade-off between complexity of the Markov chain (and thus calculations) and the average position in our evaluation task.}
\begin{tabular}{| c | c | c | c | c | c |}\cline{2-6}
\multicolumn{1}{c|}{} & \multicolumn{5}{c|}{Markov chain orders for} \\
\multicolumn{1}{c|}{} & AIC & BIC &  Significant Diff. & Prediction Task & Best Balance \\\hline 
Change-Type Paths (cf. section~\ref{sub:change type paths}) & 3 & 2 & ${_1}\eta{_3}$ & 3 & 3\\\hline
Edit-Strategy Paths (cf. section~\ref{sub:edit strategy paths}) & 4 & 3 & ${_3}\eta{_5}$ & 4 & \marked{3}\\\hline
User-Interface Sections Paths (User) (cf. section~\ref{sub:user-interface tab paths}) & 2 & 1 & ${_1}\eta{_2}$ & 1 & 1\\\hline
User-Interface Sections Paths (Concept) (cf. section~\ref{sub:user-interface tab paths}) & 2 & 1 & ${_1}\eta{_2}$ & 2 & 1\\\hline
\end{tabular}
\label{tab:results}
\end{table*}

\section{Discussion}
\label{discussion}

In section~\ref{evaluation} we have shown that the presented and adapted Markov chain model selection framework can be used to extract sequential patterns in the form of first and higher order Markov chains. 

As shown in Table~\ref{tab:results}, Markov chains of third or higher order yield the best results in our prediction tasks. The information criteria AIC and BIC, putting a negative bias on model complexity, tend to suggest minimally lower Markov chain orders. After manually inspecting and comparing the performance of the different Markov chain models, the conducted significance tests and the model complexity, we identified that a third-order Markov chain provided the best balance between said attributes for the \emph{Change-Type Paths} analysis and the \emph{Edit-Strategy Paths} analysis. For both approaches of the \emph{User-Interface Sections Paths} analyses a first-order Markov chain constitutes the best tradeoff between model complexity and performance. The identification of at least one higher-order Markov chain in our \emph{Model Selection} tasks indicates that the Markovian assumption is \emph{not} universally true for all features of the collaborative ontology-engineering change-logs. However, even if models of higher order are identified and, theoretically, provide better results than models of lower order, for the majority of the investigated change-log features a first-order Markov chain still represents the best tradeoff between model complexity and predictive power.

This result means that the previous three changes of a user contain predictive information about the change action that is most likely conducted next by that user in ICD-11. Analogously, the last change conducted by a user contains predictive information about the section of the user-interface that this user is most likely to use for the next change and if the user will stay on the same depth-level or moves up or down.

To  expand further on the usefulness of Markov chains for analyzing change-logs of collaborative ontology-engineering projects we will provide an exemplary investigation of the structure of the extracted Markov chain model for the \emph{User-Interface Paths (user-based)} analysis, including information about potential use-cases in productive environments.

\marked{
\textbf{Markov chain structure of the User-Interface Paths (user-based) Analysis:}

\begin{figure*}[!ht]
\centering
\includegraphics[width=0.7\linewidth]{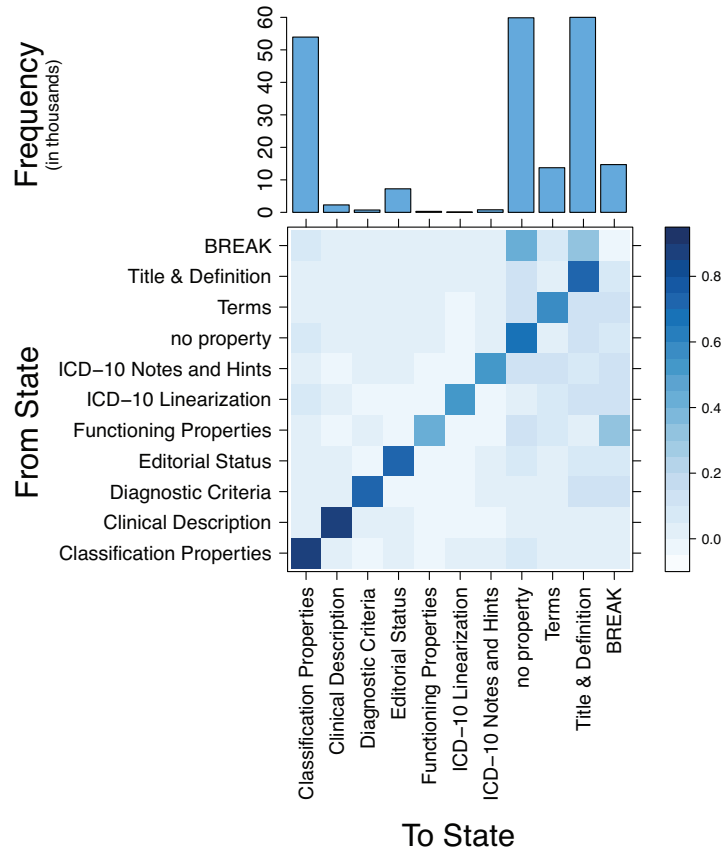}
\caption{\textbf{Results for the \emph{User-Interface Sections Paths} (user-based) analysis.}  The states for these analyses are represented by the different sections of the user-interfaces of the ontology-engineering tool iCAT (see Figure~\ref{fig:icat}). The transition probabilities for the first-order Markov chains are depicted in the transition map. Columns and rows represent the states, where rows are source states and columns are target states, indicating that a sequence always is read \emph{from row to column}. Darker colors represent higher transition probabilities while lighter colors indicate lesser transition probabilities. A clear trend towards \emph{self-loops} can be observed, meaning that changes are performed consecutively within the same sections of the user interface. The histogram depicts the absolute number of occurrences for each section for ICD-11 in alphabetical order. Sections with very low numbers of observations have been removed from the plots for reasons of readability.}
\label{fig:tus}
\end{figure*}

Figure~\ref{fig:tus} depicts the transition probabilities of a first-order Markov chain for the user-interface section sequences for properties changed by users in ICD-11. The figure clearly shows that the sections of the user-interface frequently receive consecutive changes with minimal transition probabilities to different sections of the user-interface. Note that we removed all rarely used sections from Figure~\ref{fig:tus} as they do not contain valuable information, however, their removal drastically increases the readability and ease of interpretability of Figure~\ref{fig:tus}. 

iCAT provides a special export functionality, which allows users to export parts of the ontology into a spreadsheet for quick local editing. However, no such automatic import functionality is present. To insert the edited values into the ontology, contributors have to manually add the changed properties in iCAT. This is usually done by selecting one property, changing its value and then cycling through all changed concepts where that property stays selected in the interface, allowing for easy and fast editing sessions. 

The majority of changes were concentrated on a few selected sections---\emph{Title \& Definition}, \emph{Classification Properties} and \emph{Terms}---as depicted in the histogram of Figure~\ref{fig:tus}.

Contributors to ICD-11 also exhibit a very high tendency  either to change \emph{no property} or a property of the \emph{Title \& Definition} section when resuming work after a \emph{BREAK}. \blue{The state \emph{no property} refers to all changes that do not affect the value of a property (e.g., moving a concept). Hence, these changes cannot be directly mapped to properties and sections of the user-interface. Further, the high number of \emph{no property} changes warrants further inspection in future work. In this paper, we have concentrated our analysis on properties, which can be mapped to specific parts of the user interface and provide potential actionable information for ontology-tool developers.}

\textbf{Interpretation \& Practical Implications:} When looking at the results of this analysis, we can see that the functionality of the ontology-development tool might be a deciding factor on how users interact with the ontology when contributing.
This is especially evident when considering the very high \emph{self-loop} count for ICD-11, which is most likely supported and emphasized by the export functionality present in iCAT, which allows users to export parts of the ontology into a spreadsheet, which later-on has to be manually re-inserted. Conveniently, when switching concepts, the previously selected/edited property remains selected/active in iCAT, allowing for quick edit-workflows when inserting data for the same property (and thus same section) from external resources for multiple concepts.

Furthermore, it is of no surprise that users exhibit a very high probability to consecutively change properties in the \emph{Title \& Definition} section, given that it (i) contains the most basic properties with the highest priority to be added/completed and (ii) is the default section that is displayed once a user logs into the system.

The information collected with this analysis is of potential interest for project administrators, as they can adapt the engineering process to the needs of either the community or the project itself. For example, if active collaboration for different parts of the ontology is of utmost importance, the export functionality could be restricted, only allowing an export for certain parts of the ontology. Ontology-editor developers can use the transition probabilities between different sections of the user-interface to adapt, maybe even dynamically adapt the interface towards the inherent contribution processes of the community creating the ontology in question. \blue{In particular, by further expanding the User-Interface analysis we could potentially use the results to create adaptive user interfaces that reflect and augment the personal edit-styles of contributors.} 
For example, parts of the interface could automatically adapt towards the processes of the users, relying on the transition probabilities of the extracted Markov chains, to allow for an easy transition between the current and the next, most probable, user-interface section used by a contributor. \blue{Different types of sequential paths can be used for a variety of applications. For example, we could use the chronologically ordered list of users conducting changes per class to predict which user is most likely going to change a specific class next.}
}

\section{Summary \& Conclusions}
\label{conclusion}

The detailed description of the process for applying Markov chains on the change-logs of collaborative ontology-engineering projects represents a first step towards a broader methodology to gather new insights into the ongoing processes when collaboratively engineering an ontology. 
The main contributions of this paper are as follows: (i) We provide the description of the process for applying Markov chains of varying order on collaborative ontology-engineering projects to extract and analyze sequential patterns. (ii) We categorize the types of qualitative analyses of  collaborative ontology-engineering processes that Markov chains enable us to perform. (iii) Finally, we demonstrate the usefulness of such analyses on collaborative ontology-engineering change-logs using ICD-11.

\marked{We have made the Markov chain framework publicly available\footnote{https://github.com/psinger/PathTools}, hence the only requirement for replicating the analysis for other datasets is a structured change-log of the required granularity of detail (depending on the desired analyses). Results of the same analyses may differ for different datasets, depending on a multitude of factors. For example, the used ontology editor potentially influences the way users edit the ontology (i.e., changes the edit strategy).}

In the conducted prediction experiment, several Markov chains of orders $\geq1$ have been retrieved, indicating that the Markovian assumption does not hold for all aspects of the development processes in collaborative ontology-engineering projects. To further expand on the usefulness of Markov chains, we have provided an example investigation of the structure of a first-order Markov chain and its implications and use-cases for productive environments. \marked{Note that for some of our analyses we assume the administrators and contributors to have full control over the used tools (e.g., can freely adapt, change and extend parts of the User-Interface). We are aware that this might not be the case for all collaborative ontology-engineering projects. However, we argue that the presented analyses can still provide valuable and actionable information, without having to directly edit the used tools. For example, by closely inspecting change-types and changed properties. \blue{Further, it is possible that due to restrictions in the ontology-engineering tool, users might not be able to deviate from certain paths. Hence, it is important to manually investigate and interpret the obtained patterns and avoid imposing ``one specific way``of how to use the ontology-editing tool on users.}}

\marked{For future work we plan on using the presented Markov chain analysis process to study sequential action patterns in collaborative ontology-engineering projects. As a first step, we plan on acquiring the complete change-logs for multiple ($>100$) projects created with \wpro and MoKi\footnote{https://moki.fbk.eu/website/index.php}, to analyze commonalities and differences over different collaborative ontology-engineering editors.

Further, we plan on applying the presented Markov chain analysis on these datasets to identify and investigate known and established ontology-engineering methods (e.g., HCOME, GOSPL or NeOn) and best practices ``in the wild''. 
}

As change-tracking and even click-tracking data will become available more broadly, we believe that the mapped analysis process, presented in this paper, and the potential benefits of applying Markov chains on change-logs of collaborative ontology-engineering projects, represent an important step towards even better (and simpler) ontology editors. Using sequential edit information it is possible to dynamically anticipate the editing-style of the community. Even project administrators can augment the results of the analysis, for example by allowing for easier delegation of work to the most qualified users.

We hope that the presented approach will help project administrators, ontology-engineering tool developers and, most important, the community which is developing an ontology collaboratively, to devise new approaches, tools, mechanisms or even full methodologies to increase the quality of the resulting ontology and make contributing to the projects as easy as possible.

\section*{Acknowledgements}

\small
This work was generously funded by a Marshall Plan Scholarship with support
from Graz University of Technology. Additionally, this work was partially funded by the DFG in the research project ``PoSTs II''. Further, this work is supported in part by grants GM086587 and GM103316 from U.S. National Institutes of Health.

\bibliographystyle{elsarticle-num-names}

\end{document}